# The gravitational field energy momentum tensor in the Einstein equation. Construction of the gravitational field equation

Valery B. Morozov

Email: valery_morozov@hotmail.com.

*Einstein, when he began working on the general theory of relativity, believed that energy of any kind is the source of the gravitational field. Therefore, the energy of gravity, like any energy, must be the source of the field. It was previously discovered that the energy-momentum tensor of the gravitational field is already contained in the Ricci tensor. This hypothesis is used to construct a new equation of the gravitational field.*

## 1. Introduction

Sommerfeld [1] cited the example of Lenz, which allows one to obtain a length element coinciding with the Schwarzschild solution. Of course, such constructions cannot compete with the general theory of relativity, but they can probably help to understand the connection with the special and general theories of relativity. Probably, there is nothing mysterious here, but this fact suggested the idea of refining Newton's law, bypassing the results of the general theory of relativity. It would be interesting to estimate how far elementary considerations that take into account all sources of the gravitational field, including the gravitational field itself, will allow us to advance. In the work of 1913 (Einstein A, Grossman M [2]), general requirements for solving the equations of the gravitational field were formulated. Any distribution of energy must generate a gravitational field. Including the energy of the gravitational field [2]: “one can see that, along with the components of the matter energy-tension tensor $T_{ik}$, the components of the gravitational field tensor (namely $t_{ik}$) also act as equivalent sources of the field; this requirement is obviously necessary, since the gravitational effect of the system cannot depend on the physical nature of the energy that serves as the source of the field.”

According to Einstein, the correct equation should contain the field energy-momentum tensor, for example

$$R_{\mu\nu} = \kappa\left(T_{\mu\nu} - \frac{1}{2} g_{\mu\nu} T + f_{\mu\nu} - \frac{1}{2} g_{\mu\nu} f\right). \qquad (1)$$

where $f_{\mu\nu}$ is the gravitational field energy-momentum tensor, $f = f_{\mu}^{\mu}$, $R_{\mu\nu}$ is the Ricci tensor and $\kappa = \frac{8\pi G}{c^4}$.

However, instead of the field energy tensor, Einstein obtained the pseudotensor $t_{\mu\nu}$ of the energy-momentum of the gravitational field. In addition, as it turned out later, the pseudotensor $t_{\mu\nu}$ did not contain the elements of energy density and energy flux density. In his work [3], Einstein writes:

> *"There is no reason to understand $t_4^4$ as the density of the gravitational field energy, and $(t_4^1, t_4^2, t_4^3)$ as the components of the density of the gravitational energy flux. However, one can assert the following: if the volume integral of $t_4^4$ is small compared to the volume integral of the density of the "material" energy $\mathfrak{T}_4^4$, then the right-hand side certainly represents the loss of energy by the material system."*

Thus, the hopes associated with the pseudotensor were not justified.

## 2. Incorporation of gravitational field energy into Newtonian theory

Based on the correspondence principle, Einstein's arguments about the influence of field energy on gravitation are verified by simple means [4].

The energy density of the gravitational field in the Newtonian limit is well known (see ([5] § 106, problem 1):

$$f = -\frac{(\nabla\varphi)^2}{8\pi G}, \qquad (2)$$

where $\nabla\varphi$ is the gravitational field strength.

Newton's gravity provides us with the principle of superposition and Ostrogradsky's theorem. We use this to find a correction to Newton's law of gravity. Simple calculations lead to the following result for the acceleration of fall, taking into account the field density:

$$g = -\frac{GM}{r^2}\frac{1}{1-\frac{GM}{2c^2r}}. \qquad (3)$$

On the other hand, the Schwarzschild solution leads to a similar result

$$g_s = -\frac{GM}{r^2}\frac{1}{\sqrt{1-\frac{2GM}{rc^2}}} \approx -\frac{GM}{r^2}\frac{1}{1-\frac{GM}{rc^2}}. \qquad (4)$$

A graphical comparison of these results with Newton's theory is shown in Fig. 1. A comparison of the law of attraction taking into account the energy density of the gravitational field demonstrates that this law occupies an intermediate position between the laws of Newton and Einstein.

This result is a serious argument in favor of Einstein's opinion.

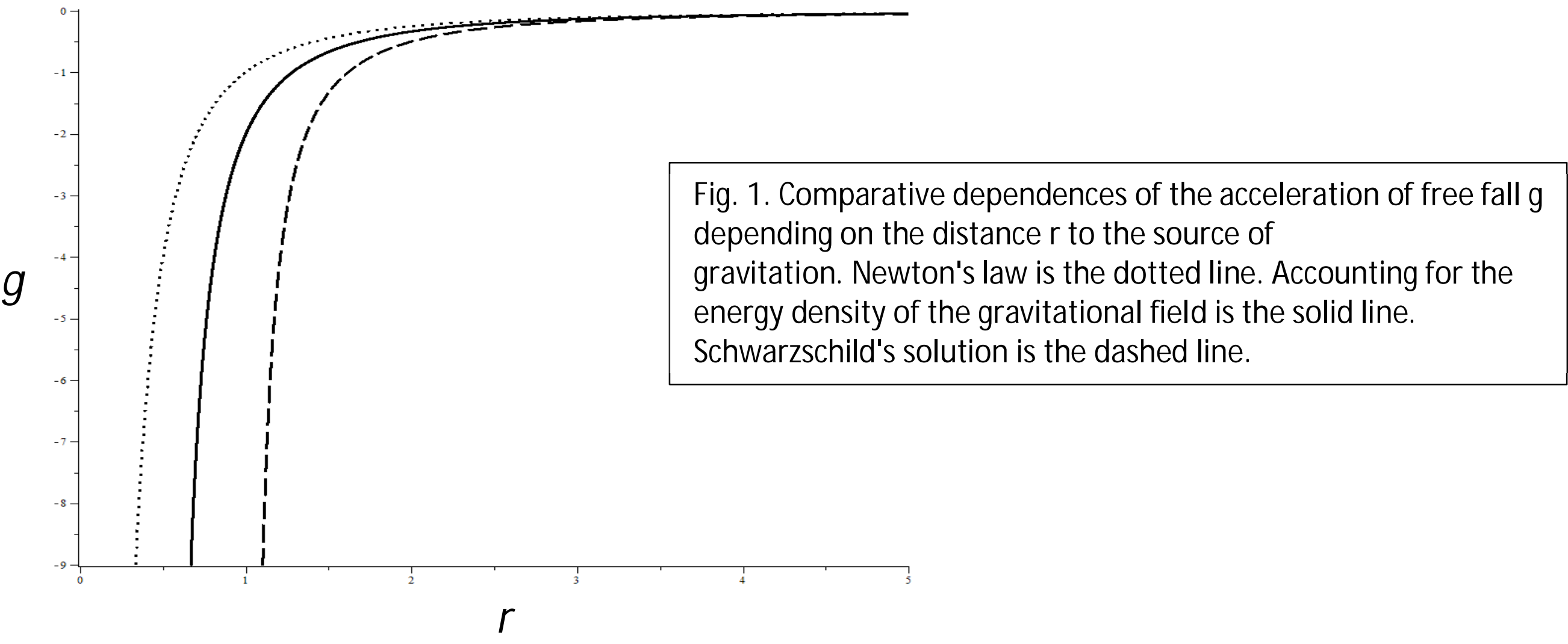

Fig. 1. Comparative dependences of the acceleration of free fall g depending on the distance r to the source of gravitation. Newton's law is the dotted line. Accounting for the energy density of the gravitational field is the solid line. Schwarzschild's solution is the dashed line.

## 3. Ricci tensor in the gravitational field equation

Following Einstein [6], we represent the Ricci tensor as the sum of two parts:

$$R_{\mu\nu} = A_{\mu\nu} + B_{\mu\nu};$$

$$A_{\mu\nu} = \frac{\partial \Gamma^{\alpha}_{\mu\nu}}{\partial x_{\alpha}} - \Gamma^{\alpha}_{\mu\beta}\Gamma^{\beta}_{\nu\alpha}; \tag{5}$$

$$B_{\mu\nu} = -\frac{\partial \Gamma^{\alpha}_{\mu\alpha}}{\partial x^{\nu}} + \Gamma^{\alpha}_{\mu\nu}\Gamma^{\beta}_{\alpha\beta} = -\frac{\partial ln\sqrt{-g}}{\partial x^{\mu}\,\partial x^{\nu}} + \Gamma^{\alpha}_{\mu\nu}\frac{\partial ln\sqrt{-g}}{\partial x^{\alpha}}.$$

However, we do not yet know whether the parts of the Ricci tensor are tensor quantities. Let us prove this.

**Theorem 1.** The simplified Christoffel symbol $\Gamma^{\alpha}_{\mu\alpha}$ is a 4-vector, $A_{\mu\nu}$ and $B_{\mu\nu}$ are tensors.

Proof. We used the formula for transforming the coordinates of the Christoffel symbols (equation (85.15), [5])

$$\Gamma^{\iota}_{\mu\alpha} = \Gamma'^{\gamma}_{\nu\xi}\frac{\partial x^{\iota}}{\partial x'^{\gamma}}\frac{\partial x'^{\nu}}{\partial x^{\mu}}\frac{\partial x'^{\xi}}{\partial x^{\alpha}} + \frac{\partial^2 x'^{\gamma}}{\partial x^{\alpha}\partial x^{\mu}}\frac{\partial x^{\iota}}{\partial x'^{\gamma}}$$

If we replace the following variables in this expression: $\iota = \alpha$ and $\gamma = \xi$, then the formula will be simplified:

$$\Gamma^{\alpha}_{\mu\alpha} = \Gamma'^{\xi}_{\nu\xi}\frac{\partial \alpha}{\partial x'^{\xi}}\frac{\partial x'^{\nu}}{\partial x^{\mu}}\frac{\partial x'^{\xi}}{\partial x^{\alpha}} + \frac{\partial^2 x'^{\xi}}{\partial x^{\alpha}\partial x^{\mu}}\frac{\partial x^{\alpha}}{\partial x'^{\xi}} = \Gamma'^{\xi}_{\nu\xi}\frac{\partial x'^{\nu}}{\partial x^{\mu}} + \frac{\partial^2}{\partial x^{\mu}\partial x'^{\xi}}x'^{\xi}$$

The last term is transformed by the theorem on differentiation of a composite function. According to the theorem on equality of mixed derivatives, the order of differentiation can be changed. Then it turns out that the last term is equal to zero, and

$$\Gamma^{\alpha}_{\mu\alpha} = \Gamma'^{\xi}_{\nu\xi}\frac{\partial x'^{\nu}}{\partial x^{\kappa}}.$$

Thus, $\Gamma^{\alpha}_{\mu\alpha}$ transforms as a covariant 4-vector and is therefore a covariant 4-vector.

Covariant differentiation of $\Gamma^{\alpha}_{\mu\alpha}$ generates the tensor

$$\frac{D\,\Gamma^{\alpha}_{\mu\alpha}}{\partial x^{\nu}} = \frac{\partial\Gamma^{\alpha}_{\mu\alpha}}{\partial x^{\nu}} - \Gamma^{\alpha}_{\mu\nu}\Gamma^{\beta}_{\alpha\beta},$$

the resulting tensor is the tensor $B_{\mu\nu}$ with the opposite sign.

From what has been proved it follows that since $R_{\mu\nu}$ is a tensor, then from $R_{\mu\nu} = A_{\mu\nu} + B_{\mu\nu}$ it follows that $A_{\mu\nu}$ is also a tensor quantity.

4. **Relationship of the $B_{\mu\nu}$ tensor with the gravitational field energy tensor**

Let us consider the tensor $B_{\mu\nu}$ in the Newtonian approximation, which is described by the stationary metric:

$$ds^2 = \left(1 + \frac{2\varphi}{c^2}\right)dt^2 - dx^2 - dy^2 - dz^2,$$

here $\varphi(x)$ is the Newtonian potential. Considering that the spatial components of the metric tensor and its determinant differ little from unity, we obtain:

$$B^{0}_{0} = \Gamma^{1}_{00}\Gamma^{0}_{10} = \frac{1}{c^2}\frac{\partial\varphi}{\partial x}\cdot\frac{1}{c^2 + 2\varphi}\frac{\partial\varphi}{\partial x} \approx \frac{1}{c^4}\left(\frac{\partial\varphi}{\partial x}\right)^2. \qquad (6)$$

The last quantity has a different sign compared to the known field energy density in the Newtonian approximation ([5] § 106, problem 1):

$$f = -\frac{(\nabla\varphi)^2}{8\pi G}. \qquad (7)$$

Equations (6) and (7) can be related through the constant $\kappa$:

$$B_{00} = \frac{8\pi G}{c^4} f_{00} = -\kappa f_{00}. \qquad (8)$$

Relation (8) allows us to relate the tensor $B_{\mu\nu}$ to the energy-momentum-stress tensor of the gravitational field $f_{\mu\nu}$:

$$B_{\mu\nu} = -\,\kappa f_{\mu\nu}, \qquad (9)$$

or

$$f_{\mu\nu} = \frac{8\pi G}{c^4}\left(\frac{\partial\Gamma^{\alpha}_{\mu\alpha}}{\partial x^{\nu}} - \Gamma^{\alpha}_{\mu\nu}\Gamma^{\beta}_{\alpha\beta}\right) = \frac{8\pi G}{c^4}\frac{D\,\Gamma^{\alpha}_{\mu\alpha}}{\partial x^{\nu}}.$$

**Theorem 2**. As the determinant of the metric tensor g approaches the value of the Minkowski tensor, which is already little different from the value of the Minkowski tensor, equation (1) approaches the Einstein equation.

Proof. As shown in equation (5), when approaching the Minkowski metric, the determinant $g \to 1$ and the tensor $f_{\mu\nu} \to 0$. This implies the asymptotic equality of equations (1) and the Einstein equation.

## 5. Gravitational field equation

Initially, Einstein assumed the inclusion of the gravitational field energy-momentum tensor in the gravitational field equation [2]. This point of view was also held by the author of these lines for quite a long time [7, 8]. However, this introduced an unjustified assumption about the sign in equation (9). The dubiousness of this assumption and the instability of the numerical solutions of the resulting equation led to the rejection of this way of constructing the field equation.

Here we use the same gravitational field energy-momentum tensor $f_{\mu\nu}$, given by equation (9), taking into account that this tensor is already contained in the equation for the principle of least action $\delta S_m + \delta S_g = 0$. Up to a constant factor, the action

$$\int \left( A_{\mu\nu} + B_{\mu\nu} - \frac{1}{2} g_{\mu\nu} R - \kappa T_{\mu\nu} \right) \delta g^{\mu\nu} \sqrt{-g} \; d\Omega = 0$$

or taking into account (9):

$$\int \left( A_{\mu\nu} - \frac{1}{2} g_{\mu\nu} R - \kappa T_{\mu\nu} - \kappa f_{\mu\nu} \right) \delta g^{\mu\nu} \sqrt{-g} \; d\Omega = 0$$

Now equation (1) is written more definitely (in mixed indices):

$$A_\mu^\nu = \kappa \left( T_\mu^\nu - \frac{1}{2} \delta_\mu^\nu T + f_\mu^\nu - \frac{1}{2} \delta_\mu^\nu f \right).$$

Returning to the Ricci tensor, we arrive at the equivalent field equation

$$R_\mu^\nu = \kappa \left( T_\mu^\nu - \frac{1}{2} \delta_\mu^\nu T - \frac{1}{2} \delta_\mu^\nu f \right). \tag{10}$$

For “empty” space (space with only a gravitational field) the field equation is:

$$R_\mu^\nu + \frac{1}{2} \delta_\mu^\nu B_\gamma^\gamma = 0;$$

or (11)

$$R_\mu^\nu + \frac{1}{2} \delta_\mu^\nu \left( \frac{\partial \Gamma_{\gamma\alpha}^\alpha}{\partial x^\gamma} - \Gamma_{\eta\eta}^\lambda \Gamma_{\lambda\beta}^\beta \right) = 0.$$

**Literature**

1. Sommerfeld A Electrodynamics Academic Press, (1952) § 38. [Зоммерфельд А. Электродинамика/ М.: Изд-во иностранной лит., (1958)]
2. Einstein A, Grossmann M Entwurf einer verallgemeinerten Relativitätstheorie und Theorie der Gravitation. Z. Math, und Phys., 1913, 62, 225—261. (Mit).
3. Einstein A. Über Gravitationwellen. Sitzungsber. preuss. Akad. Wiss., 1918, 1, 154—167.
4. Морозов В. Б. Поправка к ньютоновскому закону тяготения. Препринт. Август 2016. DOI: 10.13140/RG.2.1.4844.3121
5. Landau L D, Lifshitz E M. The Classical Theory of Fields Vol. 2 (4th ed.). Butterworth–Heinemann (1975).
6. Einstein A. Die Feldgleichungen der Gravitation. Sitzungsber. preuss. Akad. Wiss., (1915), 48, 2, 844-847.
7. Morozov V. B. Exact equation of the gravitational field based on the Einstein separation of the Ricci tensor. Parana J. Sci. Educ., v.8, n.1, (16-20), January 7, 2022.
8. Aritra Sanyal, Valery B. Morozov. About Asymptotic continuation of Einstein's equation. Preprint July 2024 DOI: 10.13140/RG.2.2.32521.04964
9. Einstein A. Uber Gravitationwellen. Sitzungsber. preuss. Akad. Wiss., 1918, 1, 154—167.